\documentclass[12pt, draftclsnofoot, onecolumn]{IEEEtran}
\usepackage{amsfonts}
\usepackage{mathrsfs}
\usepackage{cite}
\usepackage{graphicx}
\usepackage{subfigure}
\usepackage{amsmath}
\usepackage{amssymb}
\usepackage{stfloats}
\usepackage{subeqnarray}
\usepackage{cases}
\usepackage{indentfirst}
\usepackage{bm}
\usepackage{setspace}
\usepackage{url}
\usepackage{algorithmic}
\usepackage{algorithm}
\usepackage{color}
\usepackage{nomencl} 
\usepackage{parallel}
\usepackage{supertabular}
\ifCLASSINFOpdf
\else
\fi
\begin{document}
%
\title{Cost Efficiency for Economical Mobile Data Traffic Management from Users' Perspective}

\author{Jinghuan~Ma,
        Lingyang~Song,
        and~Yonghui~Li,

\thanks{
%

J.~Ma and L.~Song are with
the State Key Laboratory of Advanced Optical Communication Systems
and Networks, School of Electronics Engineering and Computer Science,
Peking University, Beijing, China, 100871, (email: \{mjhdtc, lingyang.song\}@pku.edu.cn).

Y.~Li is with School of Electrical and Information Engineering, The University of Sydney, Sydney, NSW 2006, Australia (email: \{yonghui.li\}@sydney.edu.au).

}
}
\maketitle

\begin{abstract}
Explosive demand for wireless internet services has posed critical challenges for wireless network due to its limited capacity.
To tackle this hurdle, \emph{wireless Internet service providers}~(WISPs) take the smart data pricing to manage data traffic loads.
Meanwhile, from the \emph{users' perspective}, it is also reasonable and desired to employ mobile data traffic management under the pricing policies of WISPs to improve the economic efficiency of data consumption.
In this paper we introduce a concept of cost efficiency for user's mobile data management, defined as the ratio of user's mobile data consumption benefits and its expense.
We propose an integrated cost-efficiency-based data traffic management scheme including long-term data demand planning, short-term data traffic pre-scheduling and real-time data traffic management.
The real-time data traffic management algorithm is proposed to coordinate user's data consumption to tailor to the pre-scheduled data traffic profile.
Numerical results demonstrate the effectiveness of cost efficiency framework in indicating and motivating mobile user's data consumption behavior.
The proposed management scheme can effectively motivate the user to adjust its data consumption profile to obtain the optimal data consumption cost efficiency.
\end{abstract}
\IEEEpeerreviewmaketitle

\vspace{-0.5em}
\section{Introduction}

With rapidly growing usage of smart mobile devices, the global mobile data traffic is expected to reach 30.6 exabytes per month in 2020, compared to 3.7 exabytes per month in 2015~\cite{CS-2016}.
To meet the explosive demand for mobile data, two major approaches have been adopted.
One intuitive way is to upgrade the wireless communication system for a higher network capacity~\cite{ww-2014,kc-2013}, but this will take a lengthy process after some new technology breakthroughs.
Alternatively, to support increasing data demand with current limited network capacity, network operators begin to adopt smart data pricing~\cite{as-2013,it-2012,mc-2012} as an economical incentive to affect mobile users' data consumption behavior, alleviating the network congestions.

In most existing pricing schemes typically designed from the perspective of network operators, the mobile users actually play a secondary role, as the user behavior is simply modeled by demand functions or utility functions in these frameworks~\cite{cp-2011}.
In practice, mobile users are one of the important participators of the market, whose behavior directly determines the network status~\cite{uc-2010,sl-2002,ti-2008}.
It is therefore essential to identify how to effectively reflect the users' consumption behaviors and accurately capture their consumption propensities.
To achieve this, the corresponding user-centric data traffic management tools are required to affect the users' consumption behaviors and better manage the mobile applications' data traffic profiles, while few existing scheduling works have been proposed from the angle of mobile users.

Motivated by these requirements,
in this paper we propose a metric named cost efficiency to effectively indicate the mobile user's economic efficiency of data consumption.
Cost efficiency~(CE) measures the benefit or utility of activities produced at per unit cost and it can be defined as the ratio of the obtained benefit/achieved measurable objective and the cost~\cite{AM-1945,PA-1967}.
To achieve the best efficiency, one has to optimally utilize the available resources to maximize the output/gain, so as to effectively save resources and reduce the cost.
Similarly, in mobile data management, two of the mobile user's major concerns in the data consumption are
the \emph{data consumption benefit} and \emph{the data traffic cost}.
The data consumption benefit is created by mobile applications' services~\cite{tv-2012}.
A higher priority of data service usually means a higher benefit to the user, i.e., the data service is more popular and more important to the user.
The cost of data service has a direct impact on the consumers' incentive to consume.
Therefore, the cost efficiency of data consumption can be defined as the ratio of user's data consumption benefit and its data traffic cost, expressed as
$$
{\sf CE}={{\rm Benefit} \over {\rm Cost}}.
$$
Instead of solely pursuing high consumption benefit or low cost, it is better to maximize the consumption benefit with least cost of data traffic. This will significantly improve the utilization of the expenditure on data consumption and thus save unnecessary expenses and network resources.

We further apply the cost efficiency to design a combined mobile data traffic management framework consisting of
long-term monthly data traffic planning and short-term data traffic scheduling including day-ahead pre-scheduling and real-time management.
For the long-term case, the cost efficiency as an indicator, is used to assist user in managing the daily data traffic and avoiding excessive consumption.
For the short-term case, a cost-efficiency-based data profile pre-scheduling method is proposed for the mobile user with day-ahead pricing to achieve the optimal cost efficient data profile based on the user's consumption propensity.
A real-time data traffic management algorithm is developed to coordinate mobile user's data consumption with the pre-scheduled data profile in the real-time market to maintain the cost efficiency.
Simulation results demonstrate the effectiveness of the pre-scheduling algorithm in achieving an optimal cost efficiency of the pre-scheduled data profile, and verify that
the proposed real-time data management algorithm can guarantee a relatively high cost efficiency while accommodating the uncertainty of real-time data consumption.
{
The contributions of the work are summarized below:

$\bullet$ We propose cost efficiency for the first time as an efficient metric to reflect mobile user's consumption efficiency of its data services;

$\bullet$ We analyze the cost efficiencies of data consumption in a long-term data consumption scenario with usage-based pricing to prevent inefficient mobile data consumption;

$\bullet$ We develop a cost efficient data profile pre-scheduling algorithm to optimize the mobile data allocation based on day-ahead prices;

$\bullet$ A real-time data traffic management algorithm is developed to effectively coordinate the real-time data traffic of each mobile application under the pre-scheduled data profile to maintain a high cost efficiency.
}

The rest of this paper is organized as follows. In Section~\ref{RW}, related work is presented.
In Section~\ref{SM}, we introduce the system model.
In Section~\ref{CC}, we study the case of cost efficiency in a long-term usage-based pricing market.
The cost efficient data management scheme for short-term data consumption is introduced and discussed in Section~\ref{CEM}. Simulation results and analyses are presented in Section~\ref{SIMU}, and conclusions are drawn in Section~\ref{CS}.

\vspace{-0.5em}
\section{Related Work}\label{RW}
{The economic solutions to cope with mobile data growth were proposed from either operators' perspective or users' perspective.
From the operator perspective,
}different aspects have been considered in defining pricing policies~\cite{as-2013}. In the usage-based pricing,
the pricing is directly determined by the amount of data consumption and the payment is calculated according to the amount of consumed data~\cite{pu-2011,rm-2009}.
The quality-of-service~(QoS) based pricing sets different rates for providing services with different levels of QoS~\cite{pf-2000,ep-2002}.
In~\cite{pf-2000} a QoS pricing mechanism was proposed to set prices for services with different QoS requirements, offering incentives for each consumer to choose the services that best match its needs.
The time-dependent pricing considers the dynamism of the network operation and determines the prices according to the real-time or estimated load of the network during certain periods~\cite{it-2012,On-2007,cp-2011,pu-2011,tu-2012}.
The congestion-based pricing has been proposed to dynamically increase or decrease the prices of network services/resources according to the real-time congestion level in the network, to motivate price-sensitive users to adjust their data loads~\cite{mc-2012,cp-2011,pu-2011}.
The day-ahead pricing proposed in~\cite{tu-2012}, as a variant of dynamic congestion pricing, provides pre-determined data prices
one day in advance to enable mobile subscribers to pre-schedule their data consumption profiles.

{From the user perspective,
}to model user's behavior or characteristics, the work in~\cite{rc-2002} defines a utility function of average throughput to capture user and application requirements and give the level of satisfaction for a given level of service.
The works in~\cite{dp-2005,sc-2004} use utility function of received SIR or SINR to capture the quality of users' network access links.
The works in~\cite{pu-2011,ub-2007} use utility function to indicate its satisfaction level on the achievable data rate, associated with consumption timing~\cite{pu-2011}
or user's service level agreement~\cite{ub-2007}. In~\cite{tu-2012}, user behavior is captured by its data demand volume and a waiting function representing its willingness to delay data consumption.
{The consumer attitudes in these works are captured by objectives such as maximizing utility or satisfaction. As a result, their data consumption may be affected by real-time pricing, but on the long run the consumers are still encouraged to consume in order to create more utility or satisfaction.
This may indulge consumers in immoderate and excessive data consumption.

Compared with existing works, the proposed cost efficiency is both beneficial and resource-saving, because it aims to not only maximize benefit per cost but also fully utilize the expended resource. Cost efficiency is adaptive for users with different pricing mechanisms, as we use cost efficiency to optimize mobile data consumptions in both long-term data consumption scenario and short-term scenario. Moreover, we design a model to effectively capture the time-varying consumption propensity in the managing process, ensuring a user-centric
management.
}

\vspace{-1em}
\section{System Model}\label{SM}
In this section, consuming behavioral model of user's mobile device and benefit function of data consumption are introduced for the short-term data traffic pre-scheduling
and real-time data traffic management. Pricing models for different consumption scenarios are also introduced.
To realize data traffic management of mobile device, an installed application management program is assumed to be authorized to control and record network accesses of all the applications on the mobile device.
Quality-of-service requirements of the applications will be guaranteed by the management program as long as their access requests are permitted.

\vspace{-0.5em}
\subsection{Consuming Behavioral Model}
Consider a set of $N$ active applications installed on the mobile device, denoted by $\mathcal{A}=\{1,2,\dots,a,\dots,N\}$, where each application is represented by a sequential number $a$.
Let $k$ denote the sequence of a time slot, within which the unit price of mobile data is fixed. Consequently, the data consumption amount is calculated over each time slot and
$x_k^a$ is defined as the data traffic of application $a$ in time slot $k$, which is a real number\footnote{Since the data traffic of an application in a time slot generally reaches the level of $2^{10}$ to $2^{20}$ bits, $x_k^a$ can be approximately regarded as a real number when it is counted by MB. This facilitates the theoretical analysis and allows us to use continuous optimization to approach the optimal allocation with nearly negligible errors.}.
For each $x_k^a$, we assume that it satisfies
\begin{equation}\label{con1}\small
\{x_k^a|b_k^a\le x_k^a\le B_k^a , \sum_{a\in \mathcal{A}} x_k^a \le B_k \},
\end{equation}
where $b_k^a$ and $B_k^a$ denotes the basic demand and maximum demand of application $a$ in time slot $k$, respectively.
We assume an operation cycle for the system, denoted by $\mathcal{OC}$, which consists of $K$ time slots, i.e., $\mathcal{OC}=\{1,2,\dots,k,\dots,K\}$, and use
$x^a$ to denote the demand of application $a$ in one operation cycle, defined as:
$x^a=\sum_{k \in \mathcal{OC}} x_k^a.$
The feasible set of $x^a$ is expressed as:
\begin{equation}\label{con2}\small
\{x^a|b^a\le x^a\le B^a \},
\end{equation}
where $B^a =\sum_{k\in \mathcal{OC}}B_k^a$ and $b^a$ is the minimum consumption requirement for each application in one operation cycle assigned by the user, satisfying $B^a \ge b^a\ge \sum_{k\in \mathcal{OC}}b_k^a$.
Moreover, we let $B_k$ denote the upper bound of mobile device's total data traffic in time slot $k$ to prevent excessive data consumption by the mobile device in the time slot,
satisfying
\begin{equation}\label{con3}
\sum_{a\in \mathcal{A}}x_k^a \le B_k.
\end{equation}
The allocation of all applications' data traffics in one operation cycle is denoted by vector $X$, defined as:
\begin{equation}\small
X=\{x_1^1,x_2^1,...,x_{K}^1,x_1^2,...,x_{K}^2,...,x_1^N,...,x_{K}^N\}^\mathrm{T}.
\end{equation}

\begin{figure}[!t]
\centering
\vspace{-0.3em}
\includegraphics[width=3.2in]{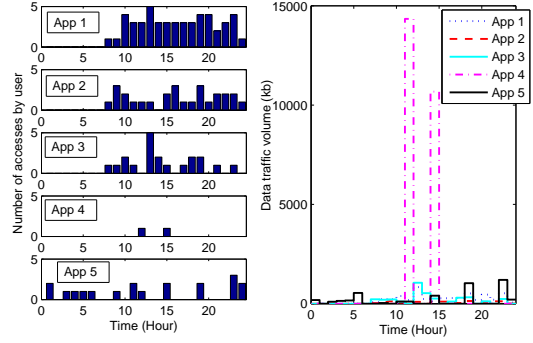}
\vspace{-0.3em}
\caption{Network accesses activated by the user.} \label{appvisit}
\vspace{-2em}
\end{figure}

To capture the consumer propensity, we record user's access frequency to each application~(i.e., how many times an application is accessed by user and runs in the foreground) in every time slot. Let $\tau_k^a$ denote the user's access frequency to application $a$ in time slot $k$.
This is because the access frequency of an application can indicate how much attention has been drawn to this service, and an application with more attention has a higher potential to please the consumer and thus deserves a higher priority in data demand scheduling. $\tau_k^a$ will be used to determine the benefit valuation parameter.
Fig.~\ref{appvisit} presents an example of 5 different applications' hourly access frequencies by the user and their hourly mobile data traffics in a day.

\subsection{Consumption Valuation Model}\label{CVP}
When people use applications for different purposes such as online social activities, reading news from the Internet, and watching online videos,
their demands and needs are satisfied through the data consumption, which can be considered as a kind of benefit.
We use utility functions~\cite{mt-1995} to model the benefits created by different applications' services.
Let $v = v(x,\omega)$ denote the benefits created over a certain amount of data consumption $x$ by an application in a time slot,
where $\omega$ is the benefit valuation parameter of an application representing its potential to create benefits that is associated with the application's access frequency.
$\omega$ also functions as a weight factor in the short-term scheduling problem that directly affects the allocation of data traffic volume among all the applications.
Let $\omega_k^a$ denote the application $a$'s potential to create benefits in time slot $k$, which is time-varying and determined based on $\tau_k^a$.
The benefit function $v_k^a$ to characterize application $a$'s contribution in time slot $k$ is given by a linear function:
\begin{equation}\label{uti}\small
v_k^a = \omega_k^a x_k^a.
\end{equation}
The linear benefit function has the following assumed properties that capture the characteristics of mobile user's consuming behavior.

\noindent\textbf{Property 1:} The benefit functions are increasing, i.e.,
\begin{equation}\label{pr1}\small
{\partial v(x_k^a,\omega_k^a)\over \partial x_k^a} > 0.
\end{equation}
This property indicates that any data consumption by the applications will create certain benefits.

\noindent\textbf{Property 2:} The benefit functions satisfy
\begin{equation}\label{pr2}\small
{\partial^2 v(x_k^a,\omega_k^a)\over \partial (x_k^a)^2} = 0.
\end{equation}
{Since we do not evaluate the benefit based on service task fulfillment, we do not differentiate the values of different bits. This property ensures the marginal benefit of an application's data consumption within a time slot is fixed.
}

\noindent\textbf{Property 3:} For a fixed $x$, a larger $\omega_k^a$ creates a larger $v(x_k^a,\omega_k^a)$, i.e.,
\begin{equation}\label{pr3}\small
{\partial v(x_k^a,\omega_k^a)\over \partial \omega_k^a} \ge 0.
\end{equation}
Since $\omega_k^a$ represents the ability to create benefits, a larger $\omega_k^a$ means more benefits created over per unit of consumed data.

\noindent\textbf{Property 4:} Benefits cannot be created without data consumption, i.e.,
\begin{equation}\label{pr4}\small
v(x_k^a=0,\omega_k^a)=0, \forall\omega_k^a>0.
\end{equation}
The setting of $\omega_k^a$ in the data profile scheduling will be addressed in Section~\ref{CEM}.

Let $V$ denote the total created benefit in one intended operation cycle, given by:
\begin{equation}\label{uti3}\small
V = \sum_{a \in \mathcal{A}} \left(\sum_{k \in \mathcal{OC}} v_k^a \right)= \sum_{a \in \mathcal{A}} \sum_{k \in \mathcal{OC}} \omega_k^a x_k^a.
\end{equation}

\subsection{Pricing Mechanisms for Data Consumption}
\subsubsection{Long-term Scenario}
Usage-based pricing has been widely applied by the mobile network operators in many countries~\cite{as-2013}. In the pricing, the operators provide data bundles with different data volumes and valid time length.
Here, we adopt the monthly data bundle plan~\cite{Singtel} that provides a fixed volume of data with a fixed price, beyond which the exceeded volume is paid at an other unit price.
Let $C(x)$ denote the monthly data traffic cost, where $x$ is the monthly total data traffic.
Hence, $C(x)$ can be expressed as:
\begin{equation}\label{ubc}\small
C(x) = \left\{ \begin{array}{ll}
C^{based} & x \le x^{limit},\\
C^{based}+ C^{unit}(x-x^{limit})& x > x^{limit},
\end{array} \right.
\end{equation}
where $C^{based}$ denotes the bundle cost, $x^{limit}$ is the capacity of the bundle and $C^{unit}$ is the unit price for exceeded part.

\begin{figure}[!t]
\centering
\vspace{-0.3em}
\subfigure{\includegraphics[width=3in]{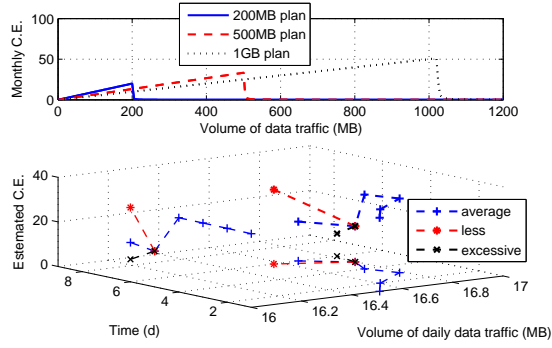}}
\caption{C.E. curves for usage-based pricing 1} \label{usagebased1}
\vspace{-0.3em}
\end{figure}
\subsubsection{Short-term Scenario}
The pricing mechanism for short-term data consumption has referred to the day-ahead pricing in~\cite{td-2011,tu-2012}, and designed as a combination of ex-ante pricing and real-time pricing. The ex-ante prices are determined by the wireless Internet service provider~(WISP) according to the observation and analysis on previous network usage, and released to the users
ahead of the next operation cycle. This has enabled and also motivated the users to pre-schedule their data traffic profiles according to the price information and pre-buy the majority of their data demands. The real-time pricing copes with the consumption uncertainty during the consuming time as it charges those who have consumed more than their planned data volume.
Noting that the time length of a time slot and the number of time slots in an operation cycle can be specified over different practical cases,
we assume similarly to~\cite{tu-2012} that a time slot equals to 1 hour and an operation cycle has 24 time slots, such that the ex-ante pricing can also be named day-ahead pricing.
In the day-ahead trading, the WISP releases the ex-ante price information
\begin{equation}\small
\{p_k| \; k=1,2,\dots,24\},
\end{equation}
where $p_k$ denotes the unit price in the $k$th hour of the next day.
Therefore, a user's pre-buying payment $C$ for the next day is given by
\begin{equation}\label{pay}\small
C=\sum_{k \in \mathcal{OC}} p_k x_k,
\end{equation}
where $x_k=\sum_{a\in \mathcal{A}}x_k^a$.
In the real-time consumption, let $p^{real}_k$ denote the real-time data unit price in hour $k$. The additional payment for the user in hour $k$ over its real real-time consumption $x'_k$
is
\begin{equation}\label{padd}\small
p^{add}_k=\max\left(p^{real}_k(x'_k-x_k),0\right).
\end{equation}
Therefore, a user's total payment for one day's data consumption is
\begin{equation}\label{pay}\small
C'=C+\sum_{k \in \mathcal{OC}} p^{add}_k.
\end{equation}

\vspace{-1em}
\section{Cost Efficiency of Long-term Data Traffic}\label{CC}
Under the usage-based pricing, if the consumer has used up the available volume before the expiry date but still want to have data service, it has to either buy an additional data bundle or consume data at a high unit price.
This is quite a common problem faced by the most of consumers as they do not control the data consumption and often use up all the data volumes in their chosen plans before the expiry date. Thus they have to spend an additional money for the additional data service, creating extra expenditure to the original budget.
Such a problem can be tackled effectively by taking the consumption cost efficiency as an indicator.

To analyze the long-term data consumption, we simplify the consumption model and assume that all variables are defined or calculated over the time length of a day.
Let $\chi_d$ denote the data traffic volume of all the applications in day $d$, and $V_d$ denote the total data consumption benefit in the day.
We have
\begin{equation}\label{uti0}\small
V_d= {\omega_d} \chi_d,
\end{equation}
where ${\omega_d}$ is the benefit valuation parameter that represents the average potential of the applications to create benefits.
As the long-term data consumption analysis focuses on the impact of daily data traffic volume $\chi_d$ on cost efficiency, we assume that ${\omega_d}$ varies slightly over days in the long-term scenario, which can be approximately equal to a constant $\bar\omega$.
\footnote{As a relative weight factor ranging from 0 to 1, $\omega$ is designed and used mainly to allocate the data traffic volume among applications in the short-term data traffic planning. There can be of great difference among the $\omega_k^a$ and $\omega^a$ in the short-term scheduling while the average benefit valuation parameter ${\omega_d}$ varies within a narrow range over days. The approximation of ${\omega_d}$ is consistent with the settings of $\omega_k^a$ and $\omega^a$ in Section~\ref{vp}.}

\begin{figure}[!t]
\centering
\subfigure{\includegraphics[width=3.25in]{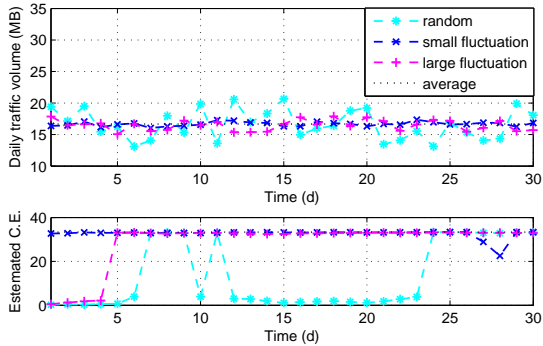}}
\vspace{-0.2em}
\caption{C.E. curves for usage-based pricing 2} \label{usagebased2}
\vspace{-1em}
\end{figure}
%

Let $\chi=\sum_{d=1}^{30} \chi_d$ denote the monthly data consumption volume.
We adopt the 30-day data bundle plans provided by Singtel~\cite{Singtel}: 200MB data within 30 days costs SGD~\$10; 500MB data within 30 days costs SGD~\$15;
1GB data within 30 days costs SGD~\$20; SGD \$0.27/10KB for the exceeded part.
Given the benefit function and pricing function, the monthly data consumption cost efficiency can be expressed as:
\begin{equation}\label{ubce}\small
{\sf CE} = \left\{ \begin{array}{ll}
{\bar\omega \chi \over C^{based} }& \chi \le x^{limit},\\
{\bar\omega \chi \over {C^{based}+ 27.648(\chi-x^{limit})}} & \chi > x^{limit}
\end{array} \right.
\end{equation}
The upper figure in Fig.~\ref{usagebased1} presents the cost efficiency versus data traffic volume for different data plans. In each plan, the cost efficiency comes to the peak at the maximum volume of a data bundle and then decreases dramatically due to the expensive retail price.
This is because when using the 200MB data bundle, the user only pays SGD~\$0.05 per 1 MB, while it has to pay SGD~\$270 for 1 MB exceeded data, which is extremely low cost effective. Therefore, it is not cost effective for users to exceed the bundle limit too much.

When we estimate the monthly cost efficiency within the expiry, $\chi$ is replaced by $\bar\chi(d)$,
which is the estimated monthly data traffic volume after day $d$, and is determined by:
$\bar\chi(d)={30 \over d}(\sum_{i=1}^d \chi_i).$
We adopt the 500MB data bundle plan in the case study. The lower figure in Fig.~\ref{usagebased1} has demonstrated the effects of different consumer behaviors on the estimated cost efficiency. After the 5th day, the estimated C.E. goes down to 16.05 from 33.32. If the consumer chooses a moderate consumption plan with 16.666MB data to consume in the 6th day, the estimated C.E. would increase a little to 17.63~(the blue line). If it chooses to consume less, i.e. 16.569MB, the estimated C.E. would increase to 33.33~(the red line).
If it chooses to consume more, i.e. 16.820MB, the estimated C.E. would decrease further to 10.09~(the black line). Fig.~\ref{usagebased2} presents the cost efficiency versus data traffic volume for different data profiles. We can see that a data profile with small fluctuation has a relative stable cost efficiency.
But it is not necessary to keep the daily data consumption volume strictly to the average level as one can cut down the data traffics for a few days if a large portion of data bundle has been consumed in the early time, so as to maintain a balance in long-term data consumption.
Moreover, fluctuation in the early days of the month has a greater impact on the estimated C.E. than that in the latter time of the month, suggesting that people had better choose a cautious consumption pattern when they just start a data plan.

\vspace{-0.5em}
\section{Cost Efficient Data Traffic Pre-scheduling and Real-time Management}\label{CEM}
\begin{figure}[!t]
\centering
\vspace{-1em}
\includegraphics[width=\linewidth]{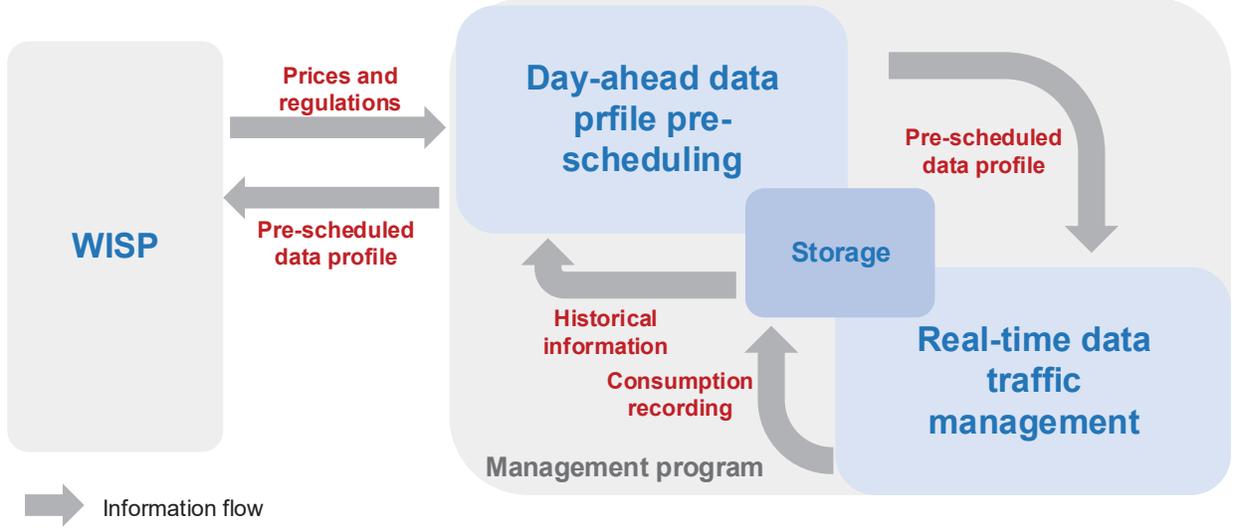}
\vspace{-0.4em}
\caption{Functioning of short-term mobile data consumption management.} \label{model}
\vspace{-2.5em}
\end{figure}

In this section, we propose a cost efficient data traffic management scheme for short-term data consumption scenario that includes day-ahead data traffic pre-scheduling and real-time data traffic management.
Fig.~\ref{model} demonstrates the functioning of data traffic management program. In the day-ahead pre-buying, the management program pre-schedules its data traffic profile according to the ex-ante price information, reports it planned data profile and pre-buys the planned data volumes for the next day.
In real-time data consumption, the management program regulates the data consumption of each application and balances the data loads of all the applications based on the pre-scheduled data profile to ensure a real-time cost efficient data consumption. The management program keeps refreshing the storage unit with the latest consumption records.

\subsection{Day-ahead Data Traffic Pre-scheduling}

\subsubsection{Problem Formulation}
In day-ahead pre-buying, the cost efficiency is defined as the ratio of the benefit of pre-bought data volume and the payment for pre-buying the data volume,
given by:
\begin{equation}\label{ce}\small
{\sf CE}={V \over C}={\sum_{a \in \mathcal{A}} \sum_{k \in \mathcal{OC}} \omega_k^a x_k^a \over \sum_{k \in \mathcal{OC}} p_k \left(\sum_{a \in \mathcal{A}} x_k^a\right)}.
\end{equation}
With the price information offered by the WISP, the management program optimizes the data profile $X$ to obtain the best cost efficiency, which can be formulated as the following objective function:
\begin{equation}\label{obj11}\small
\max_{X \in \mathcal{X}}\;\;{\sf CE}={\sum_{a \in \mathcal{A}} \sum_{k \in \mathcal{OC}} \omega_k^a x_k^a \over \sum_{k \in \mathcal{OC}} p_k \left(\sum_{a \in \mathcal{A}} x_k^a\right)},
\end{equation}
where $\mathcal{X}$ is the feasible set of all the $x_k^a$ that satisfy the consumption boundaries in (\ref{con1}), (\ref{con2}) and (\ref{con3}).
The benefit parameters $\omega_k^a$ are the weight factors for allocating data traffic volume among all the applications.
To achieve~(\ref{obj11}), the program tends to allocate more volume to the applications with higher popularity, and shift the elastic data traffic to the hours with lower prices. Moreover, if the marginal cost efficiency becomes negative due to the increase of some $x_k^a$ in their corresponding feasible ranges, the program will reduce low efficient data consumption by avoiding allocating excessive data traffic volume to $x_k^a$.
Therefore, the cost efficiency optimization can help the user to consume data smartly to reduce excessive data consumption and save money.
The optimal data traffic allocation is given by:

\begin{equation}\label{obj2}\small
\begin{aligned}
X=\;\;\;\;\;    &\arg \max_{X} {\sum_{a \in \mathcal{A}} \sum_{k \in \mathcal{OC}} \omega_k^a x_k^a \over \sum_{k \in \mathcal{OC}} p_k \left(\sum_{a \in \mathcal{A}} x_k^a\right)},\\
\text{s.t.}\;\;\;\;\; &{\rm C}1:\;b_k^a \le {x_k^a} \le B_k^a, \forall k \in \mathcal{OC},a \in \mathcal{A},\\
                &{\rm C}2:\;\sum_{a\in \mathcal{A}}x_k^a \le B_k, \forall k \in \mathcal{OC}, \\
                &{\rm C}3:\;x^a \ge b^a, \forall a \in \mathcal{A}, \\
\end{aligned}
\end{equation}
where C1 is the consumption boundaries of each application in
every time slot, C2 is the consumption limits for each
application in each time slot, and C3 is the minimum consumption requirements assigned by the user.

\subsubsection{Proposed Solution}
We adopt the fractional programs~\cite{lf-1965} to solve data traffic allocation problem. First, we prove that problem~(\ref{obj2}) is a linear fractional optimization problem, of which the objective function is a ratio of two linear continuous functions and the feasible set is continuous and convex.

\noindent\textbf{Proposition 1}: The feasible domain of problem~(\ref{obj2}) is a convex set.
\begin{proof}
The feasible domain is the intersection of three independent sets restrained by constraints C1, C2 and C3, and definitions of consumption boundaries in Section~\ref{SM} guarantee that the feasible domain is nonempty.
Let $X_{(1)}, X_{(2)} \in \mathcal{X}$, respectively, denote two different feasible data consumption profiles. We define an arbitrary real number $\alpha \in [0,1]$.
$\forall k \in \mathcal{OC}, a \in \mathcal{A}$,
$\alpha \cdot {x_{(1)k}^a} \in [\alpha \cdot b_k^a,\alpha \cdot B_k^a]$, and
$(1-\alpha) \cdot {x_{(2)k}^a} \in [(1-\alpha) \cdot b_k^a,(1-\alpha) \cdot B_k^a]$.
Therefore, $\alpha \cdot {x_{(1)k}^a} +(1-\alpha) \cdot {x_{(2)k}^a}\in [b_k^a,B_k^a].$
$\forall k \in \mathcal{OC}$,
function $(\sum_{a\in \mathcal{A}}x_k^a - B^k)$ is convex on $\mathbb{R}^N$.
$\forall a \in \mathcal{A}$,
given $\alpha \cdot {x_{(1)}^a} \ge \alpha \cdot b^a$ and
$(1-\alpha) \cdot {x_{(2)}^a} \ge (1-\alpha) \cdot b^a$,
then $\alpha \cdot {x_{(1)}^a} +(1-\alpha) \cdot {x_{(2)}^a} \ge b^a$.
Hence, the feasible domain $\mathcal{X}$ is a convex set.
\end{proof}

\noindent\textbf{Proposition 2}: Problem~(\ref{obj2}) is a linear fractional optimization problem.
\begin{proof}
The numerator in~(\ref{obj2}) is sum of linear functions as in~(\ref{uti}) differentiable over $X$.
The denominator in~(\ref{obj2}) is a linear function differentiable over $X$.
The feasible domain $\mathcal{X}$ of problem~(\ref{obj2}) is a convex set.
Thus, problem~(\ref{obj2}) is a linear fractional optimization problem.
\end{proof}

To transform problem~(\ref{obj2}) into a standard form, we construct the following vectors:
\begin{equation}\small
{\bf w}=\{\omega_1^1,\omega_2^1,...,\omega_{K}^1,\omega_1^2,...,\omega_{K}^2,...,\omega_1^N,...,\omega_{K}^N\},
\end{equation}
\begin{equation}\small
{\bf p}=\underbrace{\{p_1,p_2,...,p_K,...,p_1,p_2,...,p_K\}}_{ N \; \{p_1,p_2,...,p_K\} },
\end{equation}
\begin{equation}\small
{\bf b}=\{b_1^1,b_2^1,...,b_{K}^1,b_1^2,...,b_{K}^2,...,b_1^N,...,b_{K}^N\}^\mathrm{T},
\end{equation}
\begin{equation}\small
{\bf B}=\{B_1^1,B_2^1,...,B_{K}^1,B_1^2,...,B_{K}^2,...,B_1^N,...,B_{K}^N\}^\mathrm{T},
\end{equation}
{
where ${\bf w}$ assembles the benefit valuation parameters, ${\bf p}$ carries the price information,
${\bf b}$ collects the lower bounds and ${\bf B}$ collects the upper bounds.
}
Let $b_k=\sum_{a\in \mathcal{A}} b_k^a$ and $X^{\star}=X-{\bf b}$. We have mentioned above that $\{b^a|a\in \mathcal{A}\}$ is assigned by user and
$b^a\ge \sum_{k\in \mathcal{OC}} b_k^a,\forall a \in \mathcal{A}$.
Problem~(\ref{obj2}) can be rewritten as
\begin{equation}\label{obj5}\small
\begin{aligned}
 &\arg \max_{X^{\star}} {{\bf w}X^{\star} + {\bf wb} \over {{\bf p} X^\star + {\bf pb}} },\\
\text{s.t.} \;\;\;\;\;&{\rm C}1:\;  X^{\star} \ge 0,\\
                      &{\rm C}2:\;  AX^{\star} \le B^\star,\\
\end{aligned}
\end{equation}
in which
\begin{equation}\tiny
\begin{aligned}
& A=\\
& \left(                 
  \begin{array}{cccccc}   
    1 & 0 & 0 &...& 0 & 0\\  
    0 & 1 & 0 &...& 0 & 0\\  
    0 & 0 & 1 &0& {...} & 0\\  
    0 & {...} & {...} & {...} & {...} & 0\\  
    0 & {...} & {...} & 0 & 1 & 0\\  
    0 & {...} & {...} & 0 & 0 & 1\\  
    {[-1]_{1\times N}} & 0 & 0 & {...} &{...} & 0\\  
    {[0]_{1\times N}} & [-1]_{1\times N} & 0 & {...} & {...} & 0\\  
    {[0]_{1\times 2N}} & [-1]_{1\times N} & 0 & {...} & {...} & 0\\  
    0 & {...} & {...} & {...} & {...} & 0\\  
    0 & {...} & {...} & 0 & {[-1]_{1\times N}} & {[0]_{1\times N}} \\  
    0 & 0 & {...} & {...} & 0 & {[-1]_{1\times N}} \\  
    1 & {[0]_{1\times K-1}} & 1 &{...} & 1 &{[0]_{1\times K-1}}\\
    0 & 1 & {[0]_{1\times K-1}} & {...} & 1 & {[0]_{1\times K-2}}\\
    0 & 0 & 1 & {...} & 1 & {[0]_{1\times K-3}}\\
    0 & {...} & {...} & {...} & {...} & 0\\  
    {[0]_{1\times K-2}} & 1 & {...} & {...} & 1 & 0\\  
    {[0]_{1\times K-1}} & 1 & {...} & {...} & {[0]_{1\times K-1}} & 1\\  
  \end{array}
\right).
\end{aligned}
\end{equation}
\begin{equation}\tiny
B^\star=\left(                 
  \begin{array}{c}   
    {\bf B-b}\\
     {\sum_{k\in \mathcal{OC}} b_k^1-b^1}\\
     {\sum_{k\in \mathcal{OC}} b_k^2-b^2}\\
      {...}\\
     {\sum_{k\in \mathcal{OC}} b_k^{N-1}-b^{N-1}}\\
    {\sum_{k\in \mathcal{OC}} b_k^N-b^N}\\
         {B_1-b_1}\\
     {B_2-b_2}\\
      {...}\\
     {B_{K-1}-b_{K-1}}\\
    {B_K-b_K}\\
  \end{array}
\right).
\end{equation}
{
$A$ is a $(NK+N+K)\times(NK)$ coefficient matrix and $B^\star$ is a $(NK+N+K)\times 1$ vector including all the consumption boundaries.}
Problem~(\ref{obj5}) can be solved by the fractional programming~(FP)~\cite{fp-1983}, given in Appendix~\ref{FPintro}.
We use the Bitran-Novaes method~\cite{lp-1973}~(see Algorithm~\ref{algo} in Appendix~\ref{FPintro}) to achieve the optimal solution, which has a computational advantage.

\subsubsection{Convergence and Complexity}
To adopt the algorithm in real large-scale applications, we prove the convergence of the algorithm and study its algorithmic complexity.

\noindent\textbf{Proposition 3~(Convergence)}:~The cost efficient scheduling algorithm achieves the optimal solution within a finite number of steps.
\begin{proof}
See reference~\cite{lp-1973}.
\end{proof}
\noindent Therefore, the proposed algorithm is effective in obtaining the optimal data traffic allocation.

We assume that the proposed algorithm achieves the optimal solution in $I$ iterations.
In each iteration, obtaining the optimal ${\bf x}^*$ is a linear programming problem that can be solved by typical linear programming solvers such as
interior point method~\cite{LP-1995}.
In each inner iteration, the algorithm has to calculate the value of a function with $NK$ variables, where $N$ is the number of appliances and $K$ is the number of time slots.
The upper bound on the number of inner iteration steps is $\sqrt{NK+N+K}$, where $(NK+N+K)$ is the number of inequality constraints~\cite{LP-1995}.
Hence, the algorithmic complexity of the linear programming problem is $O(\sqrt{NK+N+K}NK)$.
The algorithmic complexity of the proposed algorithm is $O(I\sqrt{NK+N+K}KN)$.
For real large-scale applications, this algorithm can be applied in a distributed manner independently by each mobile subscriber,
and the algorithmic complexity of each single algorithm increases with the increase of $K$ and $N$. The number of iterations for convergence $I$ is verified in simulations in Section~\ref{SIMU}.

\subsection{Settings of Benefit Parameter and Consumption Constraints}\label{vp}
\subsubsection{Settings of Benefit Parameter $\omega_k^a$}
The settings of benefit parameters directly affect the allocation of data traffic volume among all the applications in a short-term data traffic pre-scheduling.
As mentioned above, an application's access frequency $\tau_k^a$ indicates its popularity and the attention drawn, which is closely related its potential to create consumption benefits.
Therefore, $\omega_k^a$ is determined based on $\tau_k^a$.
We use the latest records of $\tau_k^a$, i.e., those in the latest 24 hours, to determine the default values of $\omega_k^a$.
To simplify the presentation, in the following paragraphs, $\tau_k^a$ represents the access frequency information in the latest 24 hours.
The determination of $\omega_k^a$ follows two essential principles: an application with more attention has a higher potential to please consumer and deserves a higher priority to be granted with more data traffic volume; moreover, a time slot with higher application access frequencies indicates the period when the user has stronger incentive to consume data and thus should be assigned with more data traffic volume as well.
As $\tau_k^a$ varies in a wide range, we cannot directly use it to determine $\omega_k^a$. We propose to further express $\omega_k^a$ be determined by:
\begin{equation}\label{omega}\small
\omega_k^a=\iota^a \iota_k^a,
\end{equation}
where $\iota^a \in (0,1]$ is a weight factor indicating the relative popularity of application $a$ among all the applications,
and $\iota_k^a\in (0,1]$ is a weight factor for application $a$ to indicate its relative potential to create benefits compared with its own potentials in the other time slots.
Let $\tau^a = \sum_{k\in \mathcal{OC}} \tau_k^a$ denote the number of user's accesses to application $a$ in the intended operation cycle.
Let $\overline{\tau_k^a}=\max \{\tau_k^a|k \in \mathcal{OC} \}$ and  $\underline{\tau_k^a}=\min \{\tau_k^a|k \in \mathcal{OC} \}$.

\begin{table}[!t]
\renewcommand{\arraystretch}{1}\small
\caption{Pre-scheduling process.}
\vspace{-1em}
\label{summary}
\centering
\begin{tabular}{p{3.3in}}
\hline
$1$ User gets next day's $\{p_k^a|k\in \mathcal{OC}\}$ from the WISP.\\
$2$ Management program~(MP)'s default settings:\\
\quad $*$ calculates all $\omega_k^a$ according to (\ref{omega}).\\
\quad $*$ calculates consumption boundaries (\ref{conb}).\\
$3$ User sets $\{b^a|a\in \mathcal{A}\}$, adjusts and confirms all settings.\\
$4$ MP calculates the optimal $X$ according to (\ref{obj2}), (\ref{obj5}) and Algorithm~\ref{algo}.\\
$5$ User confirms the $X$. The MP releases demand to the WISP.\\
$6$ Contract takes effect. The MP updates the pre-scheduled data profile.\\
\hline
\vspace{-0.6em}
\end{tabular}
\vspace{-1.5em}
\end{table}

$\iota_k^a$ indicates the weight of $\tau_k^a$ in $\{\tau_k^a|k \in \mathcal{OC} \}$. In this paper, we propose $\iota_k^a$ as a function of the difference between $\tau_k^a$ and $\overline{\tau_k^a}$, and the variance of $\tau_k^a$, which is given by:
\begin{equation}\label{iotak}\small
\iota_k^a= \delta+(1-\delta){\tau_k^a \over \overline{\tau_k^a} } ^ {{(\overline{\tau_k^a}-\underline{\tau_k^a})^2/12} \over D(\tau_k^a)}.
\end{equation}
where $\delta \in (0,1)$ is the minimum value of $\iota_k^a$ to guarantee that all $\iota_k^a$ are non-zero.
This is to ensure that an application can still consume data to exchange information even when it is running on the background.
$D(\tau_k^a)$ is the variance of $\tau_k^a$, and $(\overline{\tau_k^a}-\underline{\tau_k^a})^2/12$ is the variance of an uniform distribution in $[\underline{\tau_k^a},\overline{\tau_k^a}]$
chosen as the threshold for $D(\tau_k^a)$. ${{(\overline{\tau_k^a}-\underline{\tau_k^a})^2/12} \over D(\tau_k^a)}$ is set as the power of ${\tau_k^a \over \overline{\tau_k^a} }$ to control the concavity and convexity of the function which directly effects how $\iota_k^a$ varies with ${\tau_k^a \over \overline{\tau_k^a} }$.
If $D(\tau_k^a)$ is larger than the threshold, i.e., $D(\tau_k^a)> {(\overline{\tau_k^a}-\underline{\tau_k^a})^2/12}$, indicating that all the $\tau_k^a$ are spread out around the mean and much differ from each other, function~(\ref{iotak}) becomes concave so that the difference between $\iota_k^a$ and $\iota_k^a(\overline{\tau_k^a})$ is reduced compared to the linear case $|\delta+(1-\delta){\tau_k^a \over \overline{\tau_k^a} }-1|$ when $D(\tau_k^a)= {(\overline{\tau_k^a}-\underline{\tau_k^a})^2/12}$.
If $D(\tau_k^a)< {(\overline{\tau_k^a}-\underline{\tau_k^a})^2/12}$, indicating that all the $\tau_k^a$ tend to be closer to the mean,
function~(\ref{iotak}) becomes convex so that the difference between $\iota_k^a$ and $\iota_k^a(\overline{\tau_k^a})$ in enlarged.
Similarly, let $\overline{\tau^a}=\max \{\tau^a|a \in \mathcal{A} \}$ and  $\underline{\tau^a}=\min \{\tau^a|a \in \mathcal{A} \}$.
$\iota^a$ is determined by
\begin{equation}\label{iotaa}\small
\iota^a= \delta'+(1-\delta'){\tau^a \over \overline{\tau^a} } ^ {{(\overline{\tau^a}-\underline{\tau^a})^2/12} \over D(\tau^a)},
\end{equation}
where $\delta' \in (0,1)$ is the minimum value of $\iota^a$ to guarantee that all $\iota^a$ are non-zero, $D(\tau^a)$ is the variance of $\tau^a$, $(\overline{\tau^a}-\underline{\tau^a})^2/12$ is the variance of an uniform distribution in $[\underline{\tau^a},\overline{\tau^a}]$ chosen as the threshold for $D(\tau^a)$.
We note that the default settings $\omega_k^a$ are assumed to be adjustable by user via some specific way in case the user wants to adjust its consumption plan in the next day,
while in the simulations we just use the default settings to theoretically demonstrate the performance of the proposed algorithm.

\begin{figure}[!t]
\centering
\vspace{-1em}
\includegraphics[width=\linewidth]{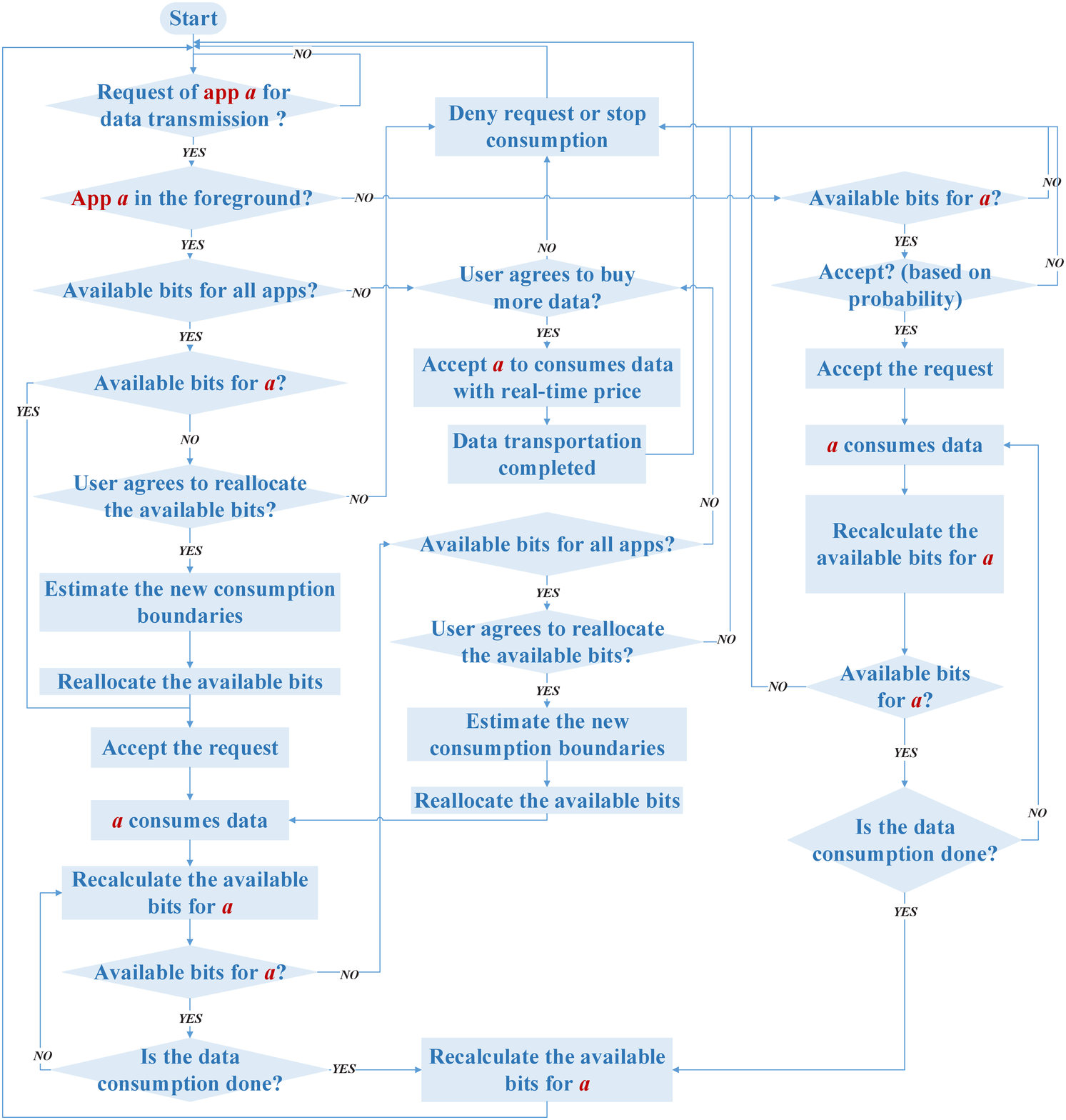}
\vspace{-0.1em}
\caption{Flow chart of the real-time data consumption management.} \label{chart}
\vspace{-1.5em}
\end{figure}

\subsubsection{Settings of Consumption Constraints}
The data traffic boundaries are determined according to user's consumption habits which can be captured from its historical consumption information.
Meanwhile, these boundaries are also assumed to be adjustable in case that the user wants to adjust its consumption plan in the next day.
To meet these requirements, the management program prepares a series of default data traffic boundaries based on user's historical consumption information.
We assume that the management program can record the data consumption information and calculate the default boundaries using the consumption records in the latest 7 days.
For example, to compute $x_k^a$ for the next day, there are seven former consumption records $\{(x_k^a)_{-1},(x_k^a)_{-2},...,(x_k^a)_{-7}\}$.
The default consumption boundaries are set as:
\begin{equation}\label{conb}\small
\begin{aligned}
b_k^a&=\min\{(x_k^a)_{-1},(x_k^a)_{-2},...,(x_k^a)_{-7}\},\\
B_k^a&=\max\{(x_k^a)_{-1},(x_k^a)_{-2},...,(x_k^a)_{-7}\},\\
B_k&=\max\{\sum_{a\in \mathcal{A}} (x_k^a)_{-1},\sum_{a\in \mathcal{A}} (x_k^a)_{-2},...,\sum_{a\in \mathcal{A}} (x_k^a)_{-7}\},\\
B^a&=\max\{(x^a)_{-1},(x^a)_{-2},...,(x^a)_{-7}\},\\
\end{aligned}
\end{equation}
{i.e., the lower bounds are set as the lowest data traffic in the latest 7 days and the upper bounds are set as the the highest data traffic in the latest 7 days.
}
Next, the user is required to provide the minimum data requirements $\{b^a|a\in \mathcal{A}\}$ of each application.

After these two steps, the user can adjust the values again if necessary and then confirm all the consumption boundaries for the optimization step.
For example, if the user does not want application $a$' to consume data in time slot $k$ it can set $B_k^a=b_k^a\equiv 0$.

With all the above settings done, the scheduling algorithm calculates the data demand and corresponding payment.
After user's confirmation on the result, the management program sends the demand information and confirmation to the WISP to make a contract.
For clear understanding, we have summarized the pre-scheduling process in Table~\ref{summary}.

\subsection{Real-time Data Traffic Management}
The real-time management program manages every application's requests of data transmission based on the pre-scheduled cost efficient data profile and benefit parameters.
The operation of real-time management in each time slot is described by the flow chart presented in Fig.~\ref{chart}.
The program uses $\{x_k^a| a\in \mathcal{A}\}$ as the original available data bits for every application. When the available bits for an application are used up, the program with a predetermined mechanism will decide whether to offer extra available bits for the application. All requests for data transmission have been divided into two kinds: requests by the foreground application have a priority over those by the background applications because the former type represents user's instant willingness to consume. Next we further introduce how the program manages these two kinds of requests.

\begin{figure}[!t]
\centering
\vspace{-0.3em}
\includegraphics[width=2.8in]{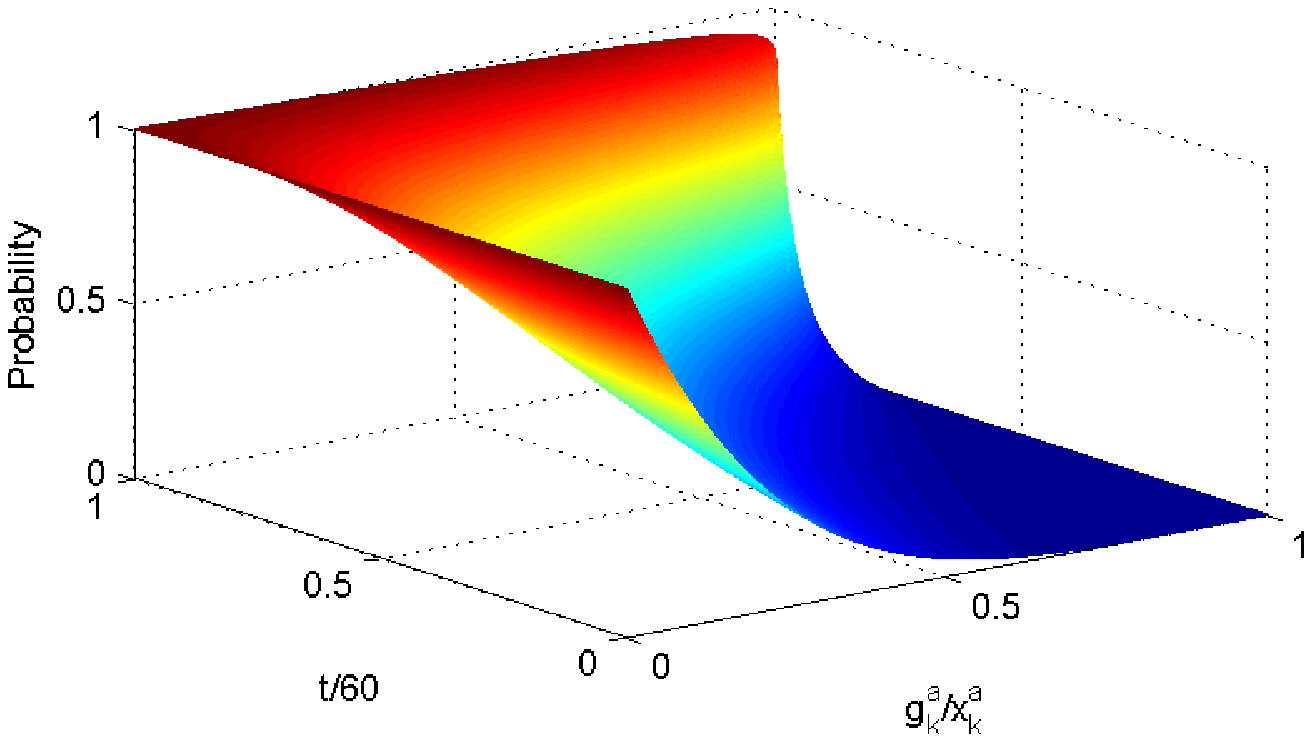}
\vspace{-0.1em}
\caption{PDF of the acceptation.} \label{pdf}
\vspace{-1em}
\end{figure}

\subsubsection{Managing background Requests}
The acceptation of request by a background application is based on a probability if there are available bits for the application.
Let $t$ denote the minutes elapsed from the beginning of the time slot and $g_k^a$ denote the consumed bits in current time slot.
The probability to accept a background request of application $a$ , denoted by $\rho_k^a$, is a function of $g_k^a$, and $t$, i.e. $\rho_k^a=\rho_k^a(g_k^a,t)$.
We assume that the probability function satisfies the following properties representing the practical considerations:

\noindent \textbf{Property 5:}
$\forall t \in [0,60]$,
\begin{equation}\small
\rho_k^a(g_k^a=0,t)=1, \rho_k^a(g_k^a=x_k^a,t)=0.
\end{equation}
That is, during each time slot, if application $a$ has not consumed any data yet, i.e., $g_k^a=0$, its request will be certainly approved.
If $a$ has consumed all its allocated data, i.e., $g_k^a=x_k^a$, its request will be certainly denied.

\noindent \textbf{Property 6:} For a fixed $t$, an increasing $g_k^a$ leads to a non-increasing $\rho_k^a(g_k^a,t)$, i.e.,
\begin{equation}\small
{\partial \rho_k^a(g_k^a,t) \over \partial g_k^a} \le 0.
\end{equation}
This is due to the fact that if $a$ has already consumed a large fraction of its allocated data, its data consumption request will be less likely to get approved in order to avoid excessive data consumption.

\noindent \textbf{Property 7:} For a fixed $g_k^a$, an increasing $t$ leads to a non-decreasing $\rho_k^a(g_k^a,t)$, i.e.,
\begin{equation}\small
{\partial \rho_k^a(g_k^a,t) \over \partial t} \ge 0,
\end{equation}
which indicates that the necessity to reserve available bits decreases with the increase of elapsed time. This is because given a fixed amount of available data, when it comes closer to the end of a time slot, the data should be more likely to be consumed so as to avoid wastage of data.

Consequently, the probability function is designed to satisfy these properties as:
\begin{equation}\label{profun}\small
\rho_k^a(g_k^a,t)=\left(1-{g_k^a \over x_k^a}\right)^{m_1 t +m_2},
\end{equation}
where $m_1 t +m_2>0$, $m_1$ and $m_2$ are two parameters that affect the shape of the function. $m_1$ and $m_2$ can be determined based on two sets of $(\rho_k^a,g_k^a,t)$.
To prevent heavy consumption of the available data in the beginning part of a time slot, at the beginning of a time slot the probability function is supposed to be convex so that $\rho_k^a$ decreases fast with the increase of $g_k^a$.
Therefore, we set $\rho_k^a(50\%x_k^a,0)$ as a reference point and assume that when $t=0$ and $g_k^a=50\%x_k^a$, the acceptation should be a small probability event, i.e., $\rho_k^a(50\%x_k^a,0)=0.05$. This means if half of the available data for $a$ has been ``dramatically'' consumed at the very beginning of a time slot, the program will probably deny the coming data consumption requests by $a$ during that period.
Nevertheless, to avoid wastage of unused data volume in a time slot, at the end of a time slot, the probability function is assumed to be concave so that $\rho_k^a$ decreases slowly and remains at a high level with the increase of $g_k^a$. Similarly, we set $\rho_k^a(90\%x_k^a,60)$ as a reference point and assume that
when $t=60$ and $g_k^a=90\%x_k^a$, the acceptation is considered as a large probability event, i.e., $\rho_k^a(90\%x_k^a,60)=0.95$.
This means if there is still $10\%$ of the available data for $a$ not being used when it comes to the end of a time slot, the program will probably accept the coming data consumption requests by $a$ during that period.
Based on the two reference points, we have $m_1={1\over 60}(\log_{0.1}0.95-\log_{0.5}0.05)$ and $m_2=\log_{0.5}0.05$. Consequently, the probability density function of the acceptation is presented in Fig.~\ref{pdf}.
When the available bits for $a$ in time slot $k$ is used up, any extra data request from the background application will be denied.

\begin{algorithm}[!t]
    \caption{Setting new consumption bounds}\label{algo2}
    \scriptsize
    \begin{algorithmic}
    \IF {$\sum_{a\in \mathcal{A}}\left({x_k^a}-{g_k^a}\right)=0$}
            \STATE Exit; (no available volume, program asks user for extra data consumption permission.)
    \ENDIF
    \STATE Set basic lower bounds ${b_k^a}'=\min\{{60g_k^a\over t},x_k^a\}, \forall a\in\mathcal{A}$;
    \STATE Set basic upper bounds ${B_k^a}'=x_k^a, \forall a\in\mathcal{A}$;
        \IF {$\sum_{a\in \mathcal{A}}\left({B_k^a}'-{b_k^a}'\right)=0$}
            \STATE  Find the application $a^*$ with the lowest $\omega_k^{a^*}$ in set $\{a|a \neq  a_1, {g_k^a}\neq {x_k^a}\}$;
            \STATE Set ${b_k^{a^*}}'={g_k^{a^*}}$;
        \ENDIF
        \STATE${B_k^{a_1}}'=$ \scriptsize$\max\{{60g_k^{a_1}\over t},g_k^{a_1}+ \sum_{a\in \mathcal{A}} \left(x_k^a-{b_k^a}'\right)\}$;
        \STATE${b_k^{a_1}}'=$ \scriptsize$\min\{{60g_k^{a_1}\over t},g_k^{a_1}+ \sum_{a\in \mathcal{A}} \left(x_k^a-{b_k^a}'\right)\}$.
\normalsize
\end{algorithmic}
\end{algorithm}

\subsubsection{Managing Foreground Requests}
Requests of data transmission by a foreground application will be directly accepted if there are available bits for the application.
When the available data for a foreground application $a$ in time slot $k$ is used up, i.e., $g_k^a=x_k^a$, the application can apply for extra available bits.
If the user permits the application for extra available bits, the management program will checks if there are available data volume for the current time slot $k$.
If so, the program will reset the lower and upper bounds of all applications' data requirements in time slot $k$, and reallocate the rest available volume for all applications by a cost efficient approach.

\begin{figure}[!t]
\centering
\vspace{-0.3em}
\includegraphics[width=3.2in]{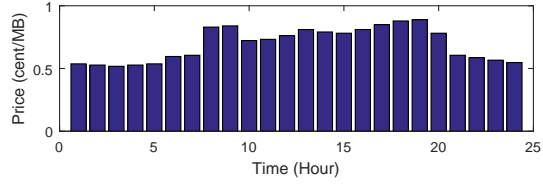}
\vspace{-0.3em}
\caption{Prices in the day-ahead market.} \label{prices}
\vspace{-1.5em}
\end{figure}

We assume that a foreground application $a_1$ has used up its available bits, i.e., $g_k^{a_1}=x_k^{a_1}$. The algorithm to reset new lower and upper bounds are presented in Algorithm~\ref{algo2}.
It guarantees that $a_1$ receives some extra data volume by increasing its lower bound beyond $g_k^{a1}$.
With the new boundaries, the cost efficient approach achieves a new optimal allocation in time slot $k$.
In order to achieve the optimal data consumption allocation, the scheduling algorithm solves the following optimization problem:
\begin{equation}\label{objx}\small
\begin{aligned}
\{{x_k^a}'|a\in \mathcal{A}\}= \;\;\;\;\;&\arg \max_{\{{x_k^a}'|a\in \mathcal{A}\}}
             {\sum_{a \in \mathcal{A}} \omega_k^a {x_k^a}' \over p_k \left(\sum_{a \in \mathcal{A}} {x_k^a}'\right)},\\
\text{s.t.}\;\;\;\;\; &{\rm C}1:\;{b_k^a}' \le {x_k^a}' \le {B_k^a}', \forall a \in \mathcal{A},\\
                      &{\rm C}2:\;\sum_{a\in \mathcal{A}} {x_k^a}'=\sum_{a\in \mathcal{A}} {x_k^a} , \\
\end{aligned}
\end{equation}
where C1 is the new consumption boundaries for all appliances, C2 guarantees that all left available bits will be reallocated.
With C2, problem~(\ref{objx}) is equal to:
\begin{equation}\label{objxx}\small
\begin{aligned}
\{{x_k^a}'|a\in \mathcal{A}\}=\;\;\;\;\;\; &\arg \max_{\{{x_k^a}'|a\in \mathcal{A}\}}
             \sum_{a \in \mathcal{A}} \omega_k^a {x_k^a}',\\
\text{s.t.}\;\;\;\;\;\; &{\rm C}1:\;{b_k^a}' \le {x_k^a}' \le {B_k^a}', \forall a \in \mathcal{A},\\
                        &{\rm C}2:\;\sum_{a\in \mathcal{A}} {x_k^a}'=\sum_{a\in \mathcal{A}} {x_k^a}, \\
\end{aligned}
\end{equation}
which can be solved by linear programming tools such as interior point method~\cite{LP-1995}.
After the cost efficient approach, the program will replace all $x_k^a$ with ${x_k^a}'$.
When there are no available bits in time slot $k$ for all the applications, the program will allow $a$ to consume data at the cost of real-time prices if the request is permitted by user,
and refresh the data traffic boundaries.

We study the algorithmic complexity of data reallocation in the real-time management.
The complexity of setting new consumption bounds is $O(N)$.
The complexity of linear programming to solve problem~(\ref{objxx}) is $O(\sqrt{2(N+1)}N)$, where $2(N+1)$ is the number of inequality constraints and $N$ is the number of applications.

\vspace{-0.5em}
\section{Simulation Results and Analyses}\label{SIMU}
In this section, we present the simulation results to demonstrate the effectiveness of the proposed load scheduling algorithm.
In addition, in day-ahead pre-scheduling we discuss how to choose a sub-optimal management strategy with limited management capability.
In the real-time data traffic management, we study the influence of the probability function of background request acceptation
on the cost efficiency of real-time data traffic management.

We first introduce the system settings in the day-ahead pre-scheduling. The hourly data traffic prices $\{p_k|k = 1,2,...,24\}$ in the day-ahead market
are set according to the parameters in~\cite{tu-2012}, which are presented in Fig.~9.
We assume there are totally $5$ applications installed in the mobile phone.
Existing works have proposed different models to generate wireless data streams~\cite{ai-2002,cu-2009,io-2014}.
To generate the historical consumption information in~(\ref{conb}), we assume that the network access request arrivals of an application is modeled by a Poisson process~\cite{ai-2002}.
Let $T_k^a$ denote the number of application $a$'s request arrivals in time slot $k$. Its probability function is
$
P(T_k^a=n) = {{e^{-\lambda^a}(\lambda^a)^n} \over {n!}},
$
where $n$ is the number of events and $\lambda^a$ is the request arrival rate of application $a$.
The expectation and variance of $T_k^a$ satisfy
$
E(T_k^a) = D(T_k^a) = \lambda^a.
$
The arrivals of user's access to an application $\tau_k^a$ are also modeled by a Poisson
process.
We have $T_k^a=\tau_k^a+\tau_k^{a*}$, where $\tau_k^{a*}$ denotes the access requests by background process of the application.
It is shown in~\cite{ai-2002} and \cite{cu-2009} that the traffic load for each network access of application such as streaming media and social application, follows a log-normal distribution. For simplicity, we employ this distribution model for all the applications and assume that in each network access, the consumed data volume follows a log-normal distribution.
Let $\epsilon^a$ denote the data consumption volume of application $a$ in a single network access.
It satisfies $\ln(\epsilon^a) \sim \rm{N}(\mu,\sigma^2)$ where $\mu$ and $\sigma^2$ are the expectation and variance of $\ln(\epsilon^a)$.
The probability density function of $\epsilon^a$ is given by
\begin{equation}\label{probab1}\small
f(\epsilon^a) = \left\{ \begin{array}{ll}
{1 \over {\sigma \epsilon^a \sqrt{2\pi}}} \exp \left[-{1 \over 2} \left( {{\ln \epsilon^a -\mu}\over \sigma} \right)^2 \right] & \epsilon^a>0,\\
{0} & \epsilon^a \le 0,
\end{array} \right.
\end{equation}
where the expectation of $\epsilon^a$ is given by
$
E(\epsilon^a) = \exp \left[ \mu +{\sigma^2 \over 2} \right],
$
and the variance of $\epsilon^a$ is
$
D(\epsilon^a) = \left[\exp  (2\mu +\sigma^2) \right]\left[\exp(\sigma^2)-1 \right].
$
The parameter settings of the data stream model are presented in Table~\ref{set}.
To determine $\omega_k^a$, we set $\delta=0.1$ and $\delta'=0.5$.

\begin{table}[!t]\scriptsize
\centering
\caption{\label{set}Settings of applications.}
\begin{tabular}{c|p{0.7cm}|p{0.7cm}|p{0.7cm}|p{0.7cm}|p{0.7cm}}
\hline
App & 1 & 2 & 3 & 4 & 5\\
\hline
$E(\tau_k^{a*})$ & [3,30] & (0,8] & (0,4] & (0,0.6] & (0,8] \\
\hline
$E(\tau_k^a)$ & [3,9] & (0,2] & (0,1] & (0,0.15] & (0,2] \\
\hline
$E(\epsilon^a)(KB)$ & 100 & 50 & 200 & 10000 & 200 \\
\hline
$D(\epsilon^a)$ & $1e^4$ & 900 & $1e^4$ & $9e^6$ & $4e^4$ \\
\hline
\end{tabular}
\vspace{-1em}
\end{table}


\begin{table}[!t]\scriptsize
\centering
\caption{\label{ind} Values of indicators for the profiles in Fig~\ref{sim1}.}
\begin{tabular}{c|c|c|c|c}
\hline
{\footnotesize Pattern}&{\footnotesize Volume(MB)} &{\footnotesize Benefit}&{\footnotesize Payment(cent)} &{\footnotesize C.E.} \\
\hline
{\footnotesize A}&45.6317&18.4074& 28.3546 & 0.6492 \\
\hline
{\footnotesize B}&45.6317&25.5956& 28.2031 & 0.9075  \\
\hline
{\footnotesize C}&41.0685&23.7148& 25.5149 & 0.9294 \\
\hline
{\footnotesize D}&45.6317&26.3196& 30.3015 & 0.8686 \\
\hline
{\footnotesize E}&49.2778&28.1991& 32.7412 & 0.8613 \\
\hline
\end{tabular}
\vspace{-2.5em}
\end{table}

In the real-time data consumption, the real-time data prices fluctuate between $100\%$ and $110\%$ of the day-ahead prices. The real-time data traffic management refers to the day-ahead scheduled profile and manages the data traffic based on the mechanism presented in Fig.~\ref{chart}. For simplicity, we assume that all foreground requests will be permitted when the
available bits are used up.
In the simulation, we assume that the numbers of foreground access requests are close to those in the previous day and the numbers of access requests by background process are about five times those by foreground applications.

\begin{figure}[!t]
\centering
\vspace{-0.3em}
\includegraphics[width=3.2in]{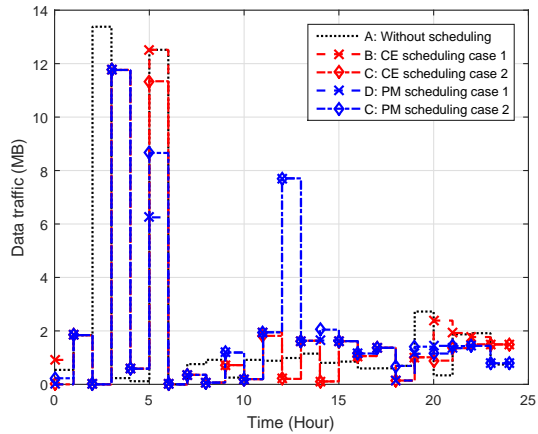}
\vspace{-0.3em}
\caption{Data consumption profiles in the day-ahead market.} \label{sim1}
\vspace{-1.5em}
\end{figure}

\vspace{-0.5em}
\subsection{Day-ahead Pre-scheduling}
We use the mentioned model to generate user's historical consumption information in the last 7 days. Based on the data traffic volumes of the applications and the arrivals of data consumption requests in the latest day, an unscheduled daily data traffic profile is generated as the baseline.
In case 1, based on the access frequencies in the latest day, we use the proposed algorithm to pre-schedule the data traffic profile under the same constraints on daily data traffic volumes of each application. Under the same constraints, we also compare the proposed algorithm with the existing profit maximization~(PM) scheme~\cite{pu-2011,rm-2009} of which the objective of the optimization can be expressed as:
\begin{equation}\label{obj1}\small
\max_{X \in \mathcal{X}}\;\;V-\eta C,
\end{equation}
where $\eta$ is the parameter that ensures a nonnegative objective and also controls the weight of cost, and is set as $0.2$.
In case 2, we slack the equality constraints on each application's daily demand and assume there is a minimum daily demand and a maximum daily demand for each application.
The minimum demand is $90\%$ of the demand in the quality constraints and the maximum demand is $110\%$ of the demand in the equality constraints.
The results are presented in Fig.~\ref{sim1} and Table~\ref{ind}.
In case 1, with the proposed scheduling on fixed data demands, the cost efficiency has improved by 39.79\% compared
to the unscheduled case, due to the increase of benefit and decrease of cost, which has demonstrated the effectiveness of the proposed algorithm in pursuing optimal cost efficiency.
Compared with the PM scheduling, the achieved benefit of CE profile is lower but close to that of the PM profile while the CE profile costs less and has a better cost efficiency.
In case 2, with relatively elastic data demands, the cost efficiency of CE profile has increased higher by 43.16\% compared to the unscheduled profile. Moreover, the data volume of CE profile has been reduced, which contributes to the reduction of data traffic load, while the PM profile has increased its data volume to obtain a better benefit and profit.
We note that since the two algorithms pursue different objectives, the comparison does not suggest the CE scheduling outperforms the PM scheduling in all aspects.

\begin{figure}[!t]
\centering
\vspace{-0.3em}
\includegraphics[width=3.2in]{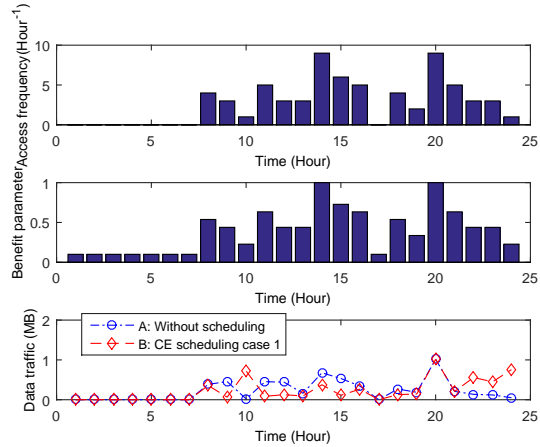}
\vspace{-0.3em}
\caption{Effect of consumer propensity.} \label{sim1.5}
\vspace{-1.5em}
\end{figure}

Fig.~\ref{sim1.5} has presented the access frequencies and benefit parameters of an application, and its corresponding data traffic profiles.
Compared with the unscheduled case, the proposed scheduling algorithm can maintain or even increase the traffics in the time slots that has high user access frequencies with relatively low prices.
However, if the price is too expensive in the time slot with high user access frequency, the algorithm tends to reduce the load or shift it to the adjacent time slots with lower prices in order to prevent the drastic drop of of cost efficiency.

\begin{figure}[!t]
\centering
\vspace{-0.3em}
\includegraphics[width=3.2in]{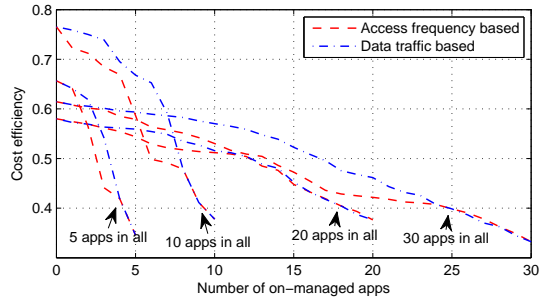}
\vspace{-0.3em}
\caption{C.E. of sub-optimal managements.} \label{sim2}
\vspace{-1.5em}
\end{figure}

\begin{figure}[!t]
\centering
\vspace{-0.3em}
\includegraphics[width=3.2in]{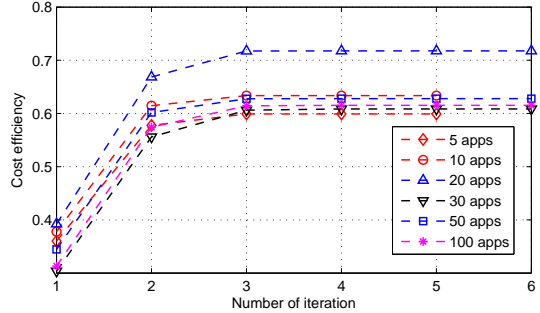}
\vspace{-0.3em}
\caption{Convergence of the algorithm.} \label{sim3}
\vspace{-1.5em}
\end{figure}

Next we consider the situation when the management program can only manage a limited number of installed applications's data traffics.
We use two different strategies to cut down the number of applications to be managed: one is to start from the application with the lowest access frequency and the other is from that with the lowest data demand. In the simulations we study the cases with different numbers of installed applications, ranging from 5 to 30.
As presented in Fig.~\ref{sim2}, in all cases, the strategy of ignoring applications with a lowest data demand performs better than that of ignoring applications with lowest access frequency.
This suggests that if a consumer only wants to manage a limited number of applications, it is suggested to manage those with relatively high data traffic demands.
As a verification of the convergence analysis, Fig.~\ref{sim3} has presented the convergence of the proposed algorithm in achieving the optimal solutions over different numbers of installed applications. It demonstrates that the performance of the proposed algorithm is efficient and stable over the increase number of installed applications.


\vspace{-0.5em}
\subsection{Real-time Data Traffic Management}
\begin{figure}[!t]
\centering
\vspace{-0.3em}
\includegraphics[width=3.2in]{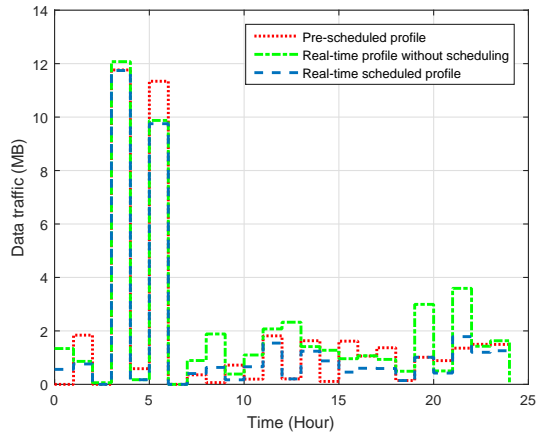}
\vspace{-0.3em}
\caption{Data traffic profiles.} \label{simR1}
\vspace{-1em}
\end{figure}

Fig.~\ref{simR1} has presented the real-time data traffic profiles compared with the pre-scheduled profile described in case 2 of Fig.~\ref{sim1}.
In the case without real-time scheduling, all access requests are permitted, while in the case with scheduling, only foreground requests will be permitted when the
available bits are used up. Both of the real-time profiles have exceeded the pre-scheduled profile in some time slots, indicating that the situation of foreground requests for data consumption with real-time data prices has been included. The cost efficiency of unscheduled profile is 0.6571 while that of the scheduled profile is 0.7141. The comparison of real-time profiles and that of the C.E. has indicated that the real-time data traffic management can reduce the unaware data traffics and maintain a relatively high cost efficiency.

\begin{figure}[!t]
\centering
\vspace{-0.3em}
\includegraphics[width=3.2in]{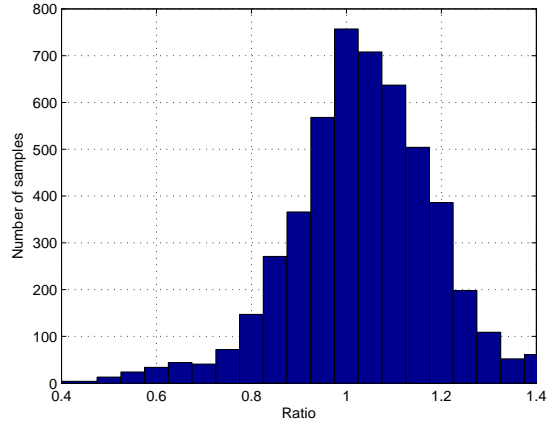}
\vspace{-0.3em}
\caption{Distribution of the ratios of cost efficiencies.} \label{simR2}
\end{figure}

We have run 5000 times of the simulations to test the effectiveness of the real-time data traffic management. Fig.~\ref{simR2} has presented the distribution of the ratio of scheduled and unscheduled data traffics. In 60.92\% of the cases the ratio are larger than 1. The cases that the ratio is smaller than 1 is due to the surplus of the pre-scheduled data traffic in real-time data consumption. As the real-time management aims to avoid excessive consumptions, it becomes less effective when the real-time data traffic tends to be far less than the pre-scheduled one.
To avoid the waste of pre-order data, the user is suggested to adjust the benefit parameters and consumption constraints for a more precise day-ahead pre-scheduling according to its real demand. Additionally, the surplus can also be caused by a strict restriction of the background access requests which is represented by the probability function~(\ref{profun}). To tackle this, a slack acceptation probability function can be used:
\begin{equation}\label{profun2}\small
\rho_k^{a'} (g_k^a,t)= \left(  \min( 1, 1- (1-\kappa)^{-1} ({g_k^a \over x_k^a}-\kappa ) ) \right)^{m_1 t +m_2},
\end{equation}
where $\kappa$ is a threshold of $g_k^a \over x_k^a$ to guarantee that if ${g_k^a \over x_k^a} \le \kappa $, $\rho_k^{a'}(g_k^a,t)=1$.
Therefore, before ${g_k^a \over x_k^a} > \kappa $, the data traffic management will accept all the background access requests.
When $\kappa=0.5$, the percentage of the cases that the ratio are larger than 1 can increase to $71.16\%$.

\section{Conclusion}\label{CS}
We proposed cost efficiency as a
metric of the economical efficiency in mobile data consumption
to optimize the consumption benefit per unit cost.
The cost efficiency can be adopted in both long-term and short-term data plans with different pricing mechanisms to
assist consumer in improving the economical efficiency of data consumption, which is verified by the case studies and the simulation results.
The effectiveness of the proposed cost efficient data traffic management for user with day-ahead and real-time pricing policy
have been demonstrated by the simulation results.
The settings of benefit parameters have effectively reflected consumer propensity in the data traffic allocation.
Besides, the cost efficiency as an economical consumption perspective, can be applied in more consumption scenarios with a flexible pricing mechanism.

\appendices
\section{Fractional Programming}\label{FPintro}
A fractional program is a nonlinear optimization problem in which the objective function is a ratio of two real functions, expressed as:
\begin{equation}\label{FPF}\small
\underset{\bm{x} \in \mathcal{D}}{\text{maximize}} \;\; f(\bm{x})={{h_1(\bm{x})} \over h_2(\bm{x})},
\end{equation}
where $\mathcal{D} \in \mathbb{R}^n$, $h_1, h_2:\mathcal{D} \rightarrow \mathbb{R}$ are both differentiable, and $h_2(\bm{x})>0$.
Studies in fractional programming mainly focus on the objective function and not on the constraint set $\mathcal{D}$, and are generally based on the assumption that $\mathcal{D}$ is a convex set of $\mathbb{R}^n$. The following discussions are also based on the assumption that $\mathcal{D}$ is a convex set.
Optimization problems in this paper belong to the concave fractional program in which the objective satisfies the following concavity/convexity assumption:
$h_1$ is concave and $h_2$ is convex on $\mathcal{D}$; $h_1$ is positive on $\mathcal{D}$ if $h_2$ is not affine.
Such nonlinear programs have the following property which is proved by~\cite{lp-1973}.

\noindent \textbf{Property 8}: In a concave fractional program~(\ref{FPF}), any local maximum is a global
maximum. If $h_1$ is strictly concave or $h_2$ is strictly convex, (\ref{FPF}) has at most one maximum.
If~(\ref{FPF}) is differentiable, then a solution of the Karush-Kuhn-Tucker conditions is a maximum of~(\ref{FPF}).

Problem~(\ref{obj2}) is a special case of the concave fractional program
where $h_1$, $h_2$ are affine functions and $\mathcal{D}$ is a convex polyhedron, and is called a linear fractional program.
The standard form of a linear fractional program can be expressed as:
\begin{equation}\label{LFP}\small
\begin{aligned}
\max_{\bf x}\;\;\;\;\;
             &L={{\bf c}^\mathrm{T}{\bf x} + {\alpha} \over {{\bf d}^\mathrm{T}{\bf x} + {\beta}} },\\
\text{s.t.} \;\;\;\;\; & A{\bf x} \le {\bf b},\\
             & {\bf x} \ge 0,\\
\end{aligned}
\end{equation}
where ${\bf c, d} \in \mathbb{R}^n$, ${\bf b} \in \mathbb{R}^m$, $\alpha, \beta \in \mathbb{R}$ and $A$ is a $m\times n$ matrix.
In addition to \textbf{Property 8}, the linear fractional programs have the property:

\noindent\textbf{Property 9}: The objective function in a linear fractional program is quasiconvex
on the feasible set $\mathcal{D}$. Therefore, a maximum is attained at a vertex of $\mathcal{D}$ if $\mathcal{D}$ is
nonempty and bounded.

To directly solve the quasiconcave program~(\ref{LFP}), we adopt the method proposed by G.~Bitran and A.~Novaes that dose not require variable transformations nor the introduction of new variables and constraints, and has a computational advantage~\cite{lp-1973}. The algorithm is as follows.

\textbf{Step 1.} Solve the problem
$$
\begin{aligned}\small
\max \;\;\;\;\;&F=<\gamma,{\bf x}>,\\
\text{s.t.}\;\;\;\;\; & A{\bf x} \le {\bf b},\\
            & {\bf x} \ge 0,\\
\end{aligned}\eqno{(A1)}\label{a1}
$$
where
$ \gamma = {\bf c}-{<{\bf c},{\bf d}>\over <{\bf d},{\bf d}>}{\bf d}$~(A2).
By the simplex routine, it achieves a suboptimal solution ${\bf x}^*$.

\textbf{Step 2.} Solve the problem
$$
\begin{aligned}\small
\max \;\;\;\;\;&F'=<[{\bf c}-L(x^*){\bf d}],{\bf x}>,\\
\text{s.t.}\;\;\;\;\; & A{\bf x} \le {\bf b},\\
            & {\bf x} \ge 0,\\
\end{aligned}\eqno{(A3)}\label{a3}
$$
where $L$ is as in~(\ref{LFP}). This yields a new feasible solution ${\bf x}^{**}$
by changing the marginal-cost line in the previous simplex routine.

\textbf{Step 3.} Compare ${\bf x}^{**}$ with ${\bf x}^*$: if ${\bf x}^{**}={\bf x}^*$, then ${\bf x}^{**}$ is the global optimum; otherwise
go to \emph{Step 2}, replacing ${\bf x}^{*}$ with ${\bf x}^{**}$ and repeating the process until ${\bf x}^*$ remains unchanged.

The first step aims to get a feasible extreme point in such a way to reduce further search.
To do this, one considers the family of hyperplanes
representing objective function~(\ref{LFP}), and chooses the hyperplane that is orthogonal
to $\beta +<{\bf d},{\bf x}>=0$. Then a simplex routine assuming an objective
function parallel to the hyperplane orthogonal to $\beta +<{\bf d},{\bf x}>=0$ is used. This hyperplane
is represented by (A1) together with (A2).
The simplex solution in the first step ${\bf x}^*$, together
with the intersection of the family of hyperplanes representing objective function~(\ref{LFP}), creates a new hyperplane that cuts the convex set in two semispaces.
In \textbf{Step 2}, the gradient of $L$ in ${\bf x}^*$ is perpendicular to the new hyperplane.
Also, every $\bf x$ that satisfies $L({\bf x}) > L({\bf x}^*)$ is located in the same semispace.
Hence, the objective function is parallel to the new hyperplane and leads to a new extreme point ${\bf x}^{**}$.
If ${\bf x}^{**}\neq {\bf x}^{*}$, then ${\bf x}^{**}$ together with the intersection of the family of
hyperplanes representing the original objective function~(\ref{LFP}), generates a new hyperplane.
The convex set is divided in two semispaces again and this will proceed until ${\bf x}^{**}= {\bf x}^{*}$.
The algorithm is summarized in Algorithm~\ref{algo}.

\begin{algorithm}[!t]\scriptsize
    \caption{The Bitran-Novaes method}\label{algo}
    \begin{algorithmic}
        \STATE \textbf{Data:} $\gamma = {\bf c}-{<{\bf c},{\bf d}>\over <{\bf d},{\bf d}>}{\bf d}$;
        \STATE Solve (A1) to obtain ${\bf x}^*$;
        \STATE ${\bf x}_0^{**}={\bf x}^*$;
        \STATE  Use $L({\bf x}^*)$ in~(A3) to obtain ${\bf x}_1^{**}$;
        \STATE $n=1$;
        \WHILE {${\bf x}_{n}^{**} \neq {\bf x}_{n-1}^{**}$}
            \STATE  Use $L({\bf x}_n^{**})$ in~(A3) to obtain ${\bf x}_{n+1}^{**}$;
            \STATE  $n++$;
        \ENDWHILE \label{code:recentEnd}
    \end{algorithmic}
\end{algorithm}


\begin{thebibliography}{1}

\bibitem{ICCC-2016}
J. Ma, L. Song, and Y. Li, ``Cost efficiency: economical mobile data traffic pre-scheduling in user's perspective,'' in \emph{Proc. IEEE/CIC ICCC 2016}, Chengdu, China, Jul. 2016.

\bibitem{CS-2016}
Cisco, ``Cisco visual networking index: Global mobile data traffic forecast, 2015-2020'', \emph{White Paper}, Feb.~2016, [Online].
Available: \url{http://www.cisco.com/c/en/us/solutions/collateral/service-provider/visual-networking-index-vni/mobile-white-paper-c11-520862.html}.


\bibitem{ww-2014}
J.~G.~Andrews \emph{et al.}, ``What will 5G be?'', \emph{IEEE J. Sel. Areas Commun.}, vol.~32, no.~6,  pp.~1065-1082, Jun.~2014.
\bibitem{kc-2013}
P.~Demestichas and A.~Georgakopoulos  ``5G on the horizon: Key challenges for the radio-access network'',  \emph{IEEE Veh. Technol. Mag.},  vol.~8,  no.~3,  pp.~47-53, Sep.~2013.

\bibitem{ot-2011}
D.-E.~Meddour, T.~Rasheed, and Y.~Gourhant, ``On the role of infrastructure sharing for mobile network operators in emerging markets,''
\emph{Comput. Netw.}, vol.~55, no.~7, pp.~1576-1591, May 2011.

%
%


\bibitem{as-2013}
S.~Sen, C.~Joe-Wong, S.~Ha, and M.~Chiang, ``A survey of smart data pricing: Past proposals, current plans, and future trends,''
\emph{ACM Computing Surveys}, vol.~46, no.~2, p.~15, Nov.~2013.

\bibitem{it-2012}
S.~Sen, C.~Joe-Wong, S.~Ha, and M.~Chiang, ``Incentivizing time-shifting
of data: A survey of time-dependent pricing for internet access,'' \emph{IEEE Comm. Magazine}, vol.~50, no.~11, pp.~91-99, Nov.~2012.

\bibitem{mc-2012}
J.~Dyaberi, et al, ``Managing cellular congestion using incentives'',
\emph{IEEE Communications Magazine}, vol.~50, no.~11, pp.~100-107, Nov.~2012.

\bibitem{cp-2011}
B.~Al-Manthari, N.~Nasser and H.~Hassanein, ``Congestion pricing in wireless cellular networks'',  \emph{IEEE Commun. Surveys Tuts.}, vol.~13, no.~3, pp.~358-371, 2011.

\bibitem{uc-2010}
Z.~Deng, Y.~Lu, K.~Wei, and Z.~Zhang, ``Understanding customer satisfaction and loyalty: An empirical study of mobile instant messages in China,'' \emph{International Journal of Information Management}, vol.~30, no.~4, pp.~289-300, Aug.~2010.

\bibitem{sl-2002}
A.~Caruana, ``Service loyalty: The effects of service quality and the mediating role of customer satisfaction'', \emph{European Journal of Marketing}, vol.~37, no.~7-8, pp.~811-828, 2002.

\bibitem{ti-2008}
H.~Chang, and S.~Chen, ``The impact of customer interface quality, satisfaction
and switching costs on e-loyalty: Internet experience as a moderator,'' \emph{Computers in Human Behavior}, vol.~24, no.~6, pp.~2927-2944, 2008.

\bibitem{AM-1945}
J.~Neumann, ``A model of general economic equilibrium,'' \emph{The Review of Economic Studies}, vol.~13, no.~1， pp.~1-9, 1945.

\bibitem{PA-1967}
D.~Chambers. ``Programming the allocation of funds subject to restrictions on reported results,'' \emph{Journal of the Operational Research Society}, vol.~18, no.~4, pp.~407-432, 1967.


\bibitem{tv-2012}
A.~Odlyzko, ``The volume and value of information,'' \emph{International Journal of Communication}, vol.~6, pp.~920-935, 2012.



\bibitem{pu-2011}
P.~Hande , M.~Chiang , R.~Calderbank and J.~Zhang, ``Pricing under constraints in access networks: Revenue maximization and congestion management'',  in \emph{Proc. IEEE INFOCOM},  pp.~1-9, 2010.

\bibitem{rm-2009}
S.~Li, J.~Huang, and S.~Li, ``Revenue maximization for communication networks with usage-based pricing,'' in \emph{Proc. IEEE GLOBECOME}, pp.~1-6, Nov.~2009.

\bibitem{pf-2000}
L.~A.~DaSilva, ``Pricing for QoS-enabled networks: A survey'', \emph{IEEE Commun. Surveys Tutorials}, vol.~3, no.~2, pp.~2-8, Apr.~2000.

\bibitem{ep-2002}
C.~Saraydar, N.~Mandayam, and D.~Goodman, ``Efficient power control via pricing in wireless data networks,'' \emph{IEEE Trans. Commun.}, vol.~50, no.~2, pp.~291-303, Feb.~2002.

\bibitem{On-2007}
H.~Shen, and T.~ Basar, ``Optimal nonlinear pricing for a monopolistic network service provider with complete and incomplete information'', \emph{IEEE J. Sel. Areas Commun.},
vol.~25, no.~6, pp.~1216-1223, Aug.~2007.

\bibitem{td-2011}
C.~Joe-Wong, S.~Ha, and M.~Chiang, ``Time-dependent broadband
pricing: feasibility and benefits,'' in \emph{Distributed Computing Systems (ICDCS), 2011 31st International Conference on}, Minneapolis, MN, Jun.~2011.

\bibitem{tu-2012}
S.~Ha, S.~Sen, C.~Joe-Wong, Y.~Im, and M.~Chiang, ``TUBE: Time-dependent pricing for mobile data,'' in \emph{Proc. of SIGCOMM'12}, pp.~13-17, Helsinki, Finland, Aug.~2012.

\bibitem{td-2014}
L.~Zhang, W.~Wu, and D.~Wang, ``Time-dependent pricing in wireless data networks: Flat-rate vs. usage-based schemes,'' in \emph{Proc. IEEE INFOCOM}, pp.~700-708, 2014.

\bibitem{rc-2002}
V.~Siris, ``Resource control for elastic traffic in cdma networks,'' in \emph{Proc. ACM International Conference on Mobile Computing and Networking
(MOBICOM)}, pp.~193-204, Atlanta, GA, Sep. 2002.


\bibitem{dp-2005}
J.~Lee, R.~Mazumdar, and N.~Shroff, ``Downlink power allocation for multi-class cdma wireless systems,''
\emph{IEEE/ACM Trans. Netw.}, vol.~13, no.~4, pp.~854-867, Aug.~2005.

\bibitem{sc-2004}
P.~Liu, P.~Zhang, S.~Jordan, and M.~Honig, ``Single-cell forward link
power allocation using pricing in wireless networks,'' \emph{IEEE Trans.
Wireless Commun.}, vol.~3, no.~2, pp.~533-543, mar.~004.

\bibitem{ub-2007}
D.~A.~Menasce, and V.~Dubey, ``Utility-based QoS brokering in service oriented architectures,'' in\emph{Proc. IEEE Int',l Conf. Web Services (ICWS ',07)}, pp.~422-430, Salt Lake City, UT, Jul.~2007.

\bibitem{mt-1995}
A.~Mas-Colell, M.~D.~Whinston, and J.~R.~Green, \emph{Microeconomic Theory,} 1st ed. New York: Oxford Univ. Press, 1995.

\bibitem{Singtel}
Data Bundles, Singapore Telecommunications Limited, 2015, [Online]. \url{http://info.singtel.com/personal/phones-plans/mobile/prepaid/data-bundles#addons_data-plans}



\bibitem{lf-1965}
K.~Swarup, ``Linear fractional functionals programming,''
\emph{Operations Research}, vol.~13, no.~6, pp.~1029-1036, Nov.~1965.

\bibitem{fp-1983}
S.~Schaible and T.~Ibaraki, ``Fractional programming,'' \emph{European J.
Operational Research}, vol.~12, no.~4, pp.~325-338, Apr.~1983.

\bibitem{lp-1973}
G.~R.~Bitran and A.~G.~Novaes, ``Linear programming with a fractional function,'' \emph{Operations Research}, vol.~21, no.~1, pp.~22-29, 1973.

\bibitem{LP-1995}
J.~Renegar, ``Linear programming, complexity theory and elementary functional analysis,'' \emph{Mathematical Programming}, vol.~70, no.~1, pp.~279-351, Oct.~1995.

\bibitem{ai-2002}
S.~Jin, A.~Bestavros, and A.~Iyengar, ``Accelerating internet streaming media delivery using network-aware partial caching'', \emph{Proceedings of IEEE International Conference on Distributed Computing Systems~(ICDCS)}, pp.~153-160, Vienna, Austria, Jul.~2002.

\bibitem{cu-2009}
F.~Benevenuto, T.~Rodrigues, M.~Cha, and V.~Almeida, ``Characterizing user behavior in online social networks'', \emph{Proceedings of the 9th ACM SIGCOMM conference on Internet measurement conference}, pp.~49-62, Chicago, IL, Nov.~2009.

\bibitem{io-2014}
L.~Zhou, Z.~Yang, H.~Wang, and M.~Guizani, ``Impact of execution time on adaptive wireless video scheduling,'' \emph{IEEE Journal on
Selected Areas in Communications}, vol.~32, no.~4, pp.~760-772, Apr.~2014.


\end{thebibliography}
\end{document}